\documentclass[aps,prb,twocolumn,groupedaddress]{revtex4}

\usepackage{graphicx}
\usepackage{dcolumn}
\usepackage{bm}
\usepackage{subfigure}

\begin{document}


\title{Ground-State and Domain-Wall Energies in the Spin-Glass Region\\
of the 2D $\pm J$ Random-Bond Ising Model}


\author{Ronald Fisch}
\email[]{ron@princeton.edu}
\affiliation{382 Willowbrook Dr., North Brunswick, NJ 08902}

\author{Alexander K. Hartmann}
\email[]{hartmann@physik.uni-goettingen.de}
\affiliation{Institut f\"{u}r Theoretische Physik, Universit\"{a}t
G\"{o}ttingen, Friedrich-Hund Platz 1, 37077 G\"{o}ttingen, Germany}


\date{\today}

\begin{abstract}
The statistics of the ground-state and domain-wall energies for the
two-dimensional random-bond Ising model on square lattices with
independent, identically distributed bonds of probability $p$ of
$J_{ij}= -1$ and $(1-p)$ of $J_{ij}= +1$ are studied. We are able to
consider large samples of up to $320^2$ spins by using sophisticated
matching algorithms. We study $L \times L$ systems, but we also
consider $L \times M$ samples, for different aspect ratios  $R = L /
M$.  We find that the scaling behavior of the ground-state energy
and its sample-to-sample fluctuations inside the spin-glass region
($p_c \le p \le 1 - p_c$) are characterized by simple scaling
functions. In particular, the fluctuations exhibit a cusp-like
singularity at $p_c$.  Inside the spin-glass region the average
domain-wall energy converges to a finite nonzero value as the sample
size becomes infinite, holding $R$ fixed.  Here, large finite-size
effects are visible, which can be explained for all $p$ by a single
exponent $\omega\approx 2/3$, provided higher-order corrections to
scaling are included.  Finally, we confirm the validity of
aspect-ratio scaling for $R \to 0$: the distribution of the
domain-wall energies converges to a Gaussian for $R \to 0$, although
the domain walls of neighboring subsystems of size $L \times L$ are
not independent.
\end{abstract}


\maketitle

\section{Introduction}

The spin glass (SG) phase\cite{EA75,reviewSG} is not stable at
finite temperature in two dimensions (2D).  When the energies are
not quantized, the behavior at $T = 0$ can be understood in terms of
a scaling theory,\cite{BM85,FH86} which is usually called the
droplet model.  This theory describes the scaling of the average
domain-wall (DW) energy $E_{dw}$ with length scale in terms of the
stiffness exponent $\theta$, i.e.
\begin{equation}
|E_{dw}| \sim L^{\theta}   \, .
\end{equation}
This exponent has a value of about $-0.28$ for 2D, almost
independent of the detailed nature of the bond distribution
\cite{AMMP03} and describes the scaling of different kinds of
excitations like domain walls and droplets, at least if large enough
system sizes are studied.\cite{droplet2003} The possibility that
quantization of the energies might lead to the special behavior
$\theta = 0$ was first pointed out by Bray and Moore.\cite{BM86} It
has become clear over the last few years that there actually is a
fundamental difference in the behavior of the 2D Ising spin glass at
zero temperature, between those cases where the energies are
quantized and those where it is not.\cite{HY01,AMMP03,WHP03}
Nevertheless, recent results \cite{JLMM06,Fis06} indicate that in
the low-temperature critical scaling regime, the behavior for
quantized and non-quantized models might be very similar.

The Hamiltonian of the Edwards-Anderson model for Ising spins is
\begin{equation}
  H = - \sum_{\langle ij \rangle} J_{ij} {S}_{i} {S}_{j}   \, ,
\label{eq:H}
\end{equation}
where each spin ${S}_{i}$ is a dynamical variable which has two
allowed states, $+1$ and $-1$.  The $\langle ij \rangle$ indicates a
sum over nearest neighbors on a simple square lattice of size $L
\times M$.  The standard model for quantized energies is the $\pm J$
model, where we choose each bond $J_{ij}$ to be an independent
identically distributed (iid) quenched random variable, with the
probability distribution
\begin{equation}
\label{eq:PJ}
  P ( J_{ij} ) = p \delta (J_{ij} + 1)~+~(1 - p) \delta (J_{ij} -
  1)   \, .
\end{equation}
Thus we actually set $J = 1$, as usual.  The concentration of
antiferromagnetic bonds is $p$, and $( 1 - p )$ is the concentration
of ferromagnetic bonds.  With the $P ( J_{ij} )$ of Eqn.
(\ref{eq:PJ}), the EA Hamiltonian is equivalent to the $Z_2$ gauge
glass model.\cite{ON93}  Wang, Harrington and Preskill\cite{WHP03}
have argued that the anomalous behavior of the $\pm J$ model is
caused by topological long-range order, as a consequence of the
gauge symmetry.

Along the $T=0$ axis this model exhibits a phase transition from a
ferromagneticly ordered phase \cite{kirkpatrick1977} for small
concentrations $p < p_c$ of the antiferromagnetic bonds to a SG
critical line at large values $p > p_c$ (and $p < 1 - p_c$ due to
the bipartite symmetry of the square lattice). Recently, this phase
transition was characterized by high-precision ground-state
calculations,\cite{WHP03,AH04} which have in particular yielded
$p_c=0.103(1)$.  The ferromagnetic phase persists at finite
temperatures $T>0$ for $p<p_c(T)$, while the spin-glass correlations
become long-range only at zero temperature.\cite{HY01,BM85,McM84b}
Interestingly, $p_c(T) > p_c(0)$,\cite{MerzChalker,WHP03,AH04} {\it
i.e.} the ferromagnetic phase is reentrant.

The first hint of the remarkable behavior of the $\pm J$ model in
the SG regime was observed by Wang and Swendsen.\cite{WS88} They
found that, although for periodic boundary conditions there is an
energy gap of $4 J$ between the ground states (GS) and the first
excited states, the specific heat when $T / J \ll 1$ appeared to be
proportional to $\exp ( - J / 2 T )$.

This result was questioned by Saul and Kardar.\cite{SK93,SK94}  The
behavior of the specific heat was finally demonstrated in a
convincing fashion by Lukic {\it et al.}\cite{LGMMR04}  Saul and
Kardar also found that the scaling of the DW entropy with size did
not appear to agree with the prediction of the droplet
model.\cite{FH88}  This issue has recently been
clarified,\cite{Fis06b} and the DW entropy scaling anomaly has been
associated with zero-energy domain walls.

 In this work we explore the behavior of the $\pm J$
model inside the full SG phase $p_c \le p \le 0.5$ (the behavior
$0.5\le p\le 1-p_c$ is equivalent due to the symmetry of the model).
In particular we study the finite-size scaling behavior of the GS
energy, of the fluctuations of the GS energy, and of the domain-wall
energy.  We also employ aspect-ratio (AR) scaling, \cite{FF69,PdN88}
{\it i.e.} we study rectangular lattices of width $L$ and height
$M$, the aspect ratio being $R = L / M$.  Carter, Bray and
Moore\cite{CBM02} have extended AR scaling to the Ising SG, and
demonstrated that studying the scaling as a function of $R$ is an
effective method for calculating the exponent $\theta$. Here, we
look at the limit $R \to 0$ and show that AR indeed works for the
$\pm J$ model as well, in contrast to previous
attempts.\cite{HBCMY02}  The entire probability distribution $P
(E_{dw})$ has a simple Gaussian form for small $R$, although, as we
will show, the underlying assumption of independent contributions to
the DW energy is not strictly valid.

\section{Methods}

We define domain walls for the SG as it was done in the seminal work
of McMillan.\cite{McM84b}  We look at differences in the GS energy
between two samples with the same set of bonds, and the same
boundary conditions in one direction, but different boundary
conditions in the other direction.  In particular we use periodic
(p) or antiperiodic (ap) boundary conditions along the {\bf x}-axis
and free boundary conditions along the {\bf y}-axis. For an example
of a DW created in such a way, cf.\ Fig.\ \ref{fig:normalDestroyed}.
For each set of bonds, the DW energy $E_{dw}$ is then defined to be
the difference in the GS energies of the two different boundary
conditions:
\begin{equation}
\label{eq:Edw}
 E_{dw} = E_{p} - E_{ap}  \, .
\end{equation}

\begin{figure}[htb]
\includegraphics[angle=0,width=0.95\columnwidth]{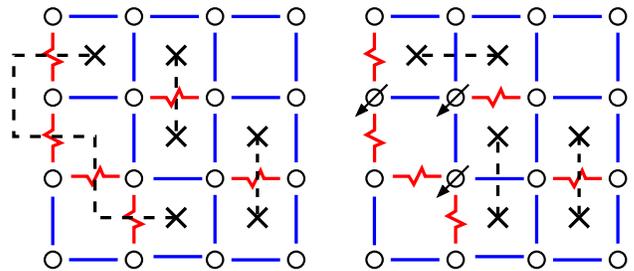}
\caption{(color online) 2D Ising spin glass with all spins up (left,
up spins not shown). Straight lines are ferromagnetic, jagged lines
are anti-ferromagnetic bonds. The dashed lines connect frustrated
plaquettes (crosses).  The bonds crossed by the dashed lines are
unsatisfied. In the right part the GS with three spins pointing down
(all others up) is shown, corresponding to the minimum number of
unsatisfied bonds.} \label{fig:matching}
\end{figure}

We use a so-called {\it matching algorithm} to calculate the
GS.\cite{PH-opt-phys2001,PH-opt-phys2004} Let us now sketch just the
basic ideas of the matching algorithm.  For the details, see
Refs.~\onlinecite{bieche1980,SG-barahona82b,derigs1991,PH-opt-phys2001}.
The algorithm allows us to find ground states for lattices which are
planar graphs.  This is the reason why we apply (anti-)periodic
boundary conditions only in one direction, {\bf x}, while the other
direction, {\bf y}, has free boundary conditions. In the left part
of Fig.~\ref{fig:matching} a small 2D system with (for simplicity)
free boundary conditions in both directions is shown. All spins are
assumed to be ``up'', hence all antiferromagnetic bonds are not
satisfied. If one draws a dashed line perpendicular to each broken
bond, one ends up with the situation shown in the figure: all dashed
lines start or end at frustrated plaquettes and each frustrated
plaquette is connected to exactly one other frustrated plaquette by
a dashed line. Each pair of plaquettes is then said to be {\it
matched}.  In general, closed loops of broken bonds unrelated to
frustrated plaquettes can also appear, but this is possible only for
excited states.  Now, one can consider the frustrated plaquettes as
the vertices and all possible pairs of connections as the edges of a
(dual) graph.  The dashed lines are selected from the edges
connecting the vertices and called a {\it perfect} matching, since
{\it all} plaquettes are matched. One can assign weights to the
edges in the dual graph, the weights are equal to the sum of the
absolute values of the bonds crossed by the dashed lines. The weight
$\Lambda$ of the matching is defined as the sum of the weights of
the edges contained in the matching. As we have seen, $\Lambda$
counts the broken bonds, hence, the energy of the configuration is
given by
\begin{equation}
 E = -\sum_{\langle i,j \rangle} |J_{ij}| + 2\Lambda \, ,
\end{equation}
with $|J_{ij}| = 1$ from Eqn. (\ref{eq:PJ}).  Note that this holds
for {\it any} configuration of the spins, if one also includes
closed loops in $\Lambda$, since a corresponding matching always
exists.

Obtaining a GS means minimizing the total weight of the broken bonds
(see right panel of Fig.~\ref{fig:matching}). This automatically
forbids closed loops of broken bonds, so one is looking for a {\it
minimum-weight perfect matching}. This problem is solvable in
polynomial time.  The algorithms for minimum-weight perfect
matchings\cite{MATCH-cook,MATCH-korte2000} are among the most
complicated algorithms for polynomial problems. Fortunately the LEDA
library offers a very efficient implementation,\cite{PRA-leda1999}
which we have applied here.

With free boundary conditions in at least one direction, $E_{dw}$ is
a multiple of 2 (since $J = 1$). When periodic or antiperiodic
boundary conditions are applied in both the {\bf x} and {\bf y}
directions, however, $E_{dw}$ becomes a multiple of 4 if the number
of $-1$ bonds is even.  If the number of $-1$ bonds is odd, $E_{dw}$
takes on values $(4 n + 2)$ with these boundary conditions. When $M$
is odd, changing the boundary conditions in the {\bf x} direction
from periodic to antiperiodic changes the number of $-1$ bonds from
odd to even, or vice versa.

The behavior of the DW energy for the $\pm J$ model with
$R > 1$ was discussed previously.\cite{HBCMY02}  For the boundary
conditions we are using, the average $\langle |E_{dw}| \rangle$
goes to zero exponentially as $R$ becomes much greater than one.
Here $\langle ... \rangle$ denotes an average over the ensemble of
random bond distributions for an $L \times M$ lattice. We can
understand this result by thinking of the system as consisting of
blocks of size $M \times M$ pasted together along the {\bf x}
direction.  The probability of having a zero-energy domain wall in
each block is almost independent of the other blocks.  The same
thing happens when $M$ is even if the boundary conditions in the
{\bf y} direction are periodic or antiperiodic. However, if $M$ is
odd and the boundary conditions in the {\bf y} direction are
periodic or antiperiodic, $\langle |E_{dw}| \rangle$ goes to 2 at
large $R$, because then $E_{dw} = 0$ is not allowed. Since a
critical exponent should be independent of boundary conditions,
this is a(nother) demonstration that $\theta = 0$ for the $\pm J$
model in 2D.

As pointed out earlier\cite{HBCMY02}, these results for large $R$
do not agree with the scaling law prediction of Carter, Bray and
Moore,\cite{CBM02} due to the special role of zero-energy domain
walls.  On the other hand, in the limit $R \to 0$ with our
boundary conditions, the prediction of Carter, Bray and Moore is
\begin{equation}
\label{eq:AR}
 \langle | E_{dw} | \rangle \sim L^{\theta} ( M / L )^{(d-1)/2}
, ~ ~ M \gg L    \, .
\end{equation}
This limit has not been studied before for the $\pm J$ model. When
$\theta = 0$, $\langle |E_{dw}| \rangle$ should scale as $R^{-1/2}$
in 2D. For small $R$ we may think of the lattice as consisting of $L
\times L$-sized subsystems stacked in the {\bf y} direction, with
$E_{dw}$ being the sum of the DW energies of the subsystems.
Therefore, by the central limit theorem, we anticipate that the
probability distribution of $E_{dw}$ should approach a Gaussian
distribution in the limit of small $R$.  In the spin-glass region of
the phase diagram, the center of this limiting Gaussian will
approach zero as $L$ increases.  We will show that our numerical
results for the small $R$ limit are indeed in good agreement with
these expectations, even though the subsystems are not fully
independent.

\section{Numerical Results}

\subsection{Ground-state energy}

We begin the study of the $T=0$ behavior of the $\pm J$ random bond
model by studying the GS energy for different concentrations $p$ of
the antiferromagnetic bonds [$p=0.05$ (ferromagnetic phase),
$p=0.11, 0.12, 0.13, 0.14, 0.15, 0.17, 0.2, 0.25, 0.3, 0.35, 0.4,
0.45$ and $p=0.5$] and different system  sizes $L=4, \ldots, 320$
for square samples $N = L \times L$, {\it i.e.} aspect ratio $R=1$.
All results are averages over many different realizations of the
disorder.  The minimum number of independent samples used varies
with size, ranging between typically 100000 ($L=4$) to typically
10000 ($L=320$).

\begin{figure}[ht]
\includegraphics[width=0.9\columnwidth]{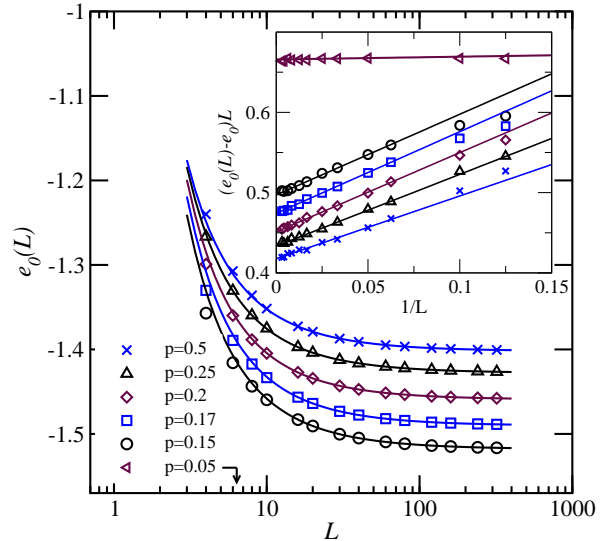}
\caption{\label{fig:E0}(color online) The symbols show the average
ground state energy $e_0(L)$ for the square systems ($R=1$) as
function of the system size $L$ for selected values of $p$.  The
lines give the results to the fits ($L\ge 16$) (see text). The data
for $p=0.05$ look similar, but is outside the frame ($e_0=-1.8025$).
The inset shows the same data, rescaled as $(e_0(L)-e_0) \times L$
so that the expected behavior yields straight lines when plotted as
function of $1 / L$. The data for $p = 0.05$ have been shifted
downwards by 0.2 for better visibility.}
\end{figure}

In Fig.\ \ref{fig:E0}, the GS energy per spin $e_0(L)$ is shown as a
function of the system size for selected values of $p$. This
finite-size scaling behavior yields particular insight into the
physics and seems to be related \cite{bouchaud2003,CHK04} to the
stiffness exponent $\theta$.  It has been argued \cite{CHK04} that
for the case of open boundary conditions in exactly one direction
and periodic boundary conditions in the other direction the GS
energy per spin follows to lowest orders the form
\begin{equation}
\label{eq:e0L}
  e_0(L) = e_0 + \frac{b}{L} + \frac{c}{L^2} + \frac{d}{L^{2-\theta}}  \; .
\end{equation}
Note that the arguments used in Ref.\ \onlinecite{CHK04} should
apply for the ferromagnetic phase $p \le 0.103$ as well. Since we
expect $\theta = 0$ everywhere in the SG phase (see below) and
$\theta = 1$ for the ferromagnetic phase, we can restrict ourselves
to the contributions  $e_0$, $b / L$ and $c / L^2$. When
fitting\cite{gnuplot} the data to this $e_0(L)$, fits with very high
quality result, as shown by lines in Fig.\ \ref{fig:E0}. This is
confirmed when plotting the data and the fitting functions in the
form $(e_0(L)-e_0)L$ as a function of $1/L$.  This should give,
according to Eqn.~(\ref{eq:e0L}), a linear behavior with slope $c$
and ordinate intersection $b$. We see that indeed the data follows
the functional form well, although even higher order corrections
seem to be present. They become significant at very small sizes, but
we cannot quantify them within the statistical accuracy of the data.
We only observe that this correction term seems to have an opposite
sign for small ($p_c \le p \le 0.25$) compared to larger values of
$p$. Furthermore, the behavior of the $c / L^2$ term is quite
similar inside the SG phase for different values of $p$, while the
behavior in the ferromagnetic phase $(p=0.05)$ is very different.
The value of the ordinate intercept $b$ is monotonic in $p$, as
expected.\cite{CHK04}

\begin{figure}
\includegraphics[width=0.9\columnwidth]{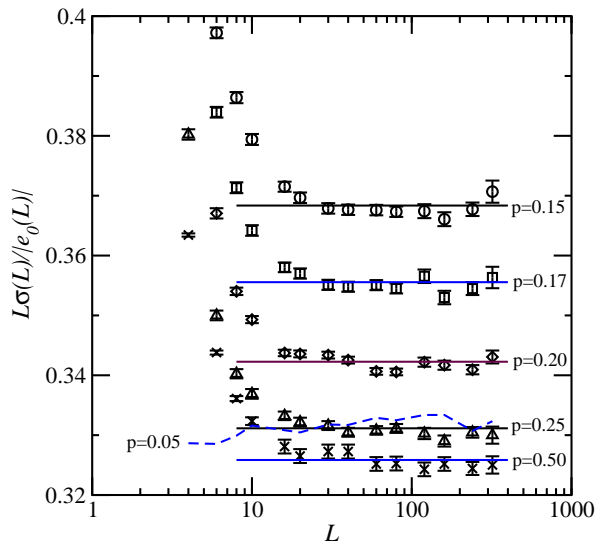}
\caption{\label{fig:widthPE0}(color online) The symbols show the
width $\sigma$ of the distributions of the GS energy $e_0$ for
square systems ($R=1$) as function of system size $L$ for selected
values of $p\ge p_c$. The data is rescaled with the average GS
energy $e_0(L)$ and with the leading behavior $L$. Additionally, the
dashed line displays the result for $p=0.05$. The full lines show
fits to constants at larger system sizes.}
\end{figure}

Next, we consider the width $\sigma$ of the distribution of GS
energies per spin. For short-range $d$-dimensional spin glasses, it
has been proven \cite{wehr1990} that $\sigma \sim L^{-d/2}$, i.e.
$\sigma \sim L^{-1}$ in this case. To remove finite-size effects
caused by our special choice of boundary conditions, we normalize
$\sigma$ by the finite-size GS energy. Thus, we plot $L \sigma /
|e_0(L)|$ as a function of system size and expect a horizontal line
at $c_\sigma(p) = \lim_{L \to \infty} L \sigma / e_0(L)$. As seen in
Fig.\ \ref{fig:widthPE0}, this is indeed the case.  Only corrections
for very small system-sizes become visible. We have also included
the data for $p=0.05$, which behave in the same way. Hence, here the
ferromagnetic phase and the SG phase cannot be distinguished just
from the asymptotic behavior, in contrast to the behavior of the
mean alone.

\begin{figure}
\includegraphics[width=0.9\columnwidth]{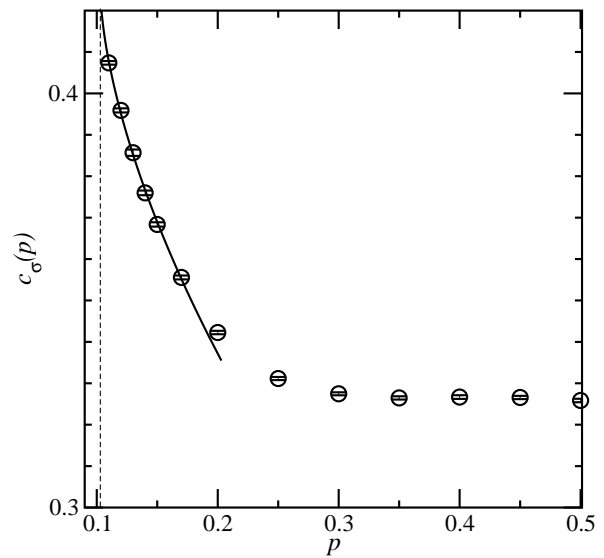}
\caption{\label{fig:widthConst} The symbols show the limiting value
$c_\sigma(p)=\lim_{L\to\infty} L\sigma/e_0(L)$ for the width
$\sigma$ of the distributions of the GS energy (as shown in Fig.\
\ref{fig:widthPE0}) as a function of $p$. The solid line represents
a fit according to Eqn.~(\ref{eq:fitCsigma}), see text.}
\end{figure}

On the other hand, as visible in Fig.\ \ref{fig:widthPE0}, the value
of $c_\sigma(p)$ increases when approaching the phase transition
$p\to p_c$. For a more detailed analysis, we have plotted
$c_\sigma(p)$ as function of $p$ inside the SG phase in
Fig.~\ref{fig:widthConst}, also for values of $p$ closer to $p_c$
than those shown in Fig.\ \ref{fig:widthPE0}. One can see a strong
increase when approaching the critical concentration. Fitting a
power law (with $p_c=0.103$)
\begin{equation}
c_\sigma(p) = C_c+K(p-p_c)^{\kappa} \label{eq:fitCsigma}
\end{equation}
in the range $p\in[0.1\ldots 0.17]$ yields values $C_c=0.422(3)$,
$K=-0.38(4)$ and $\kappa=0.64(5)$, {\it i.e.} a cusp at the phase
transition. Note that the value of the scaling exponent is very
close to the exponent $\omega=2/3$, which describes the finite-size
scaling of the DW energy $E_{dw}$; see below.

Nevertheless, the basic observation of the (almost) universal
behavior inside the SG region, which we find when looking at the GS
energy, becomes even more apparent,  when studying the behavior of
the DW energy, which we do in the next section.

\subsection{Domain-wall energy}

\begin{figure}
\includegraphics[width=0.9\columnwidth]{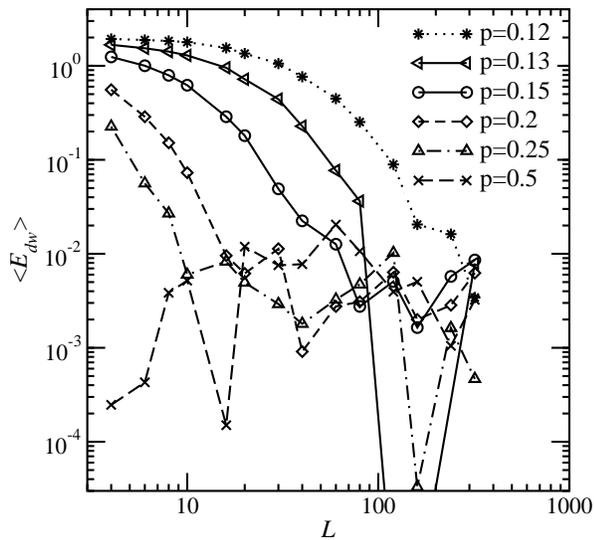}
\caption{\label{fig:Emean} The average value of the DW energy as
function of system size for selected values of $p$. Error bars are
very small for values $\langle E_{dw} \rangle > 0.1$ but relatively
large where the DW energy is small. Thus, they are omitted for
readability. Note the double-logarithmic axes.  Lines are guides to
the eye.}
\end{figure}

For all samples considered in the previous section, we have also
calculated the DW energy as defined in Eqn.~(\ref{eq:Edw}).  In
Fig.\ \ref{fig:Emean} the average value $\langle E_{dw}\rangle$ of
the DW energy is shown as a function of system size for selected
values of $p$.  Close to the phase transition $p_c=0.103$, the
systems exhibit a high degree of ferromagnetic order for small
sizes, leading to relatively large values of the average DW energy.
This explains why very large system sizes were needed in Ref.\
\onlinecite{AH04} to determined the location $p_c=0.103$ of the
phase transition precisely. For larger values $p\ge 0.2$, the
ferromagnetic correlation length is small.

\begin{figure}
\includegraphics[width=0.9\columnwidth]{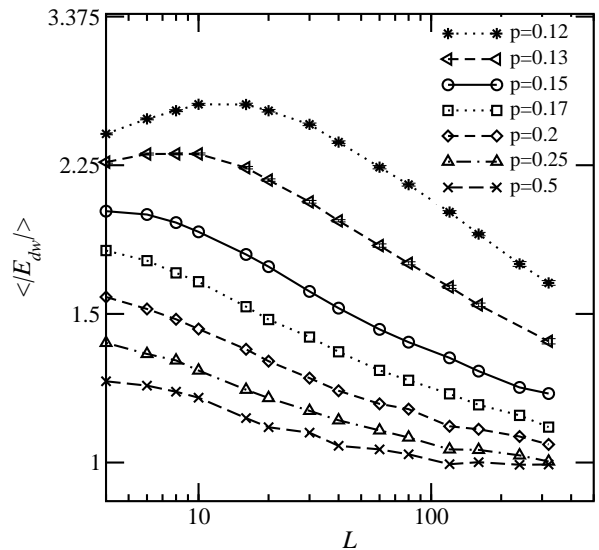}
\caption{\label{fig:Estiff} The average absolute value of the DW
energy as a function of system size for selected values of $p$. Note
the double-logarithmic axes. Lines are guides to the eye.}
\end{figure}

In Fig.\ \ref{fig:Estiff} the average absolute value $\langle
|E_{dw}| \rangle$ of the DW energy is shown as a function of system
size for selected values of $p$. For very small values of $p$, close
to the phase transition, this DW energy  increases with system size
for small system sizes. Hence, if only slow algorithm were
available, the system would look like exhibiting an ordered phase
({\it i.e.} $\theta > 0$). When going to larger system sizes, it
becomes apparent that this is only a finite-size effect. For
intermediate values of $p$, the DW energy decreases with growing
system size, but no saturation of  $\langle |E_{dw}| \rangle$ is
visible on the accessible length scales, hence the data look similar
to the results for Gaussian systems \cite{HY01} with $\theta < 0$.
When looking at the results for $p > 0.25$, it becomes apparent that
$\langle |E_{dw}| \rangle$ converges to a finite value. This is due
to the quantized nature \cite{AMMP03} of the possible values for
$E_{dw}$. Therefore, it is clear that also for intermediate values
of $p$, the DW energy converges to a finite plateau-value as well,
but for much larger system sizes. Hence, the model provides a
striking example that finite-size corrections can persist up to huge
length scales. Note that the convergence to a non-zero plateau value
is compatible with the results from the previous section, where we
have found that inside the SG phase the scaling function
Eqn.~(\ref{eq:e0L}) with $\theta = 0$ describes the data well
everywhere.

\begin{figure}
\includegraphics[width=0.9\columnwidth]{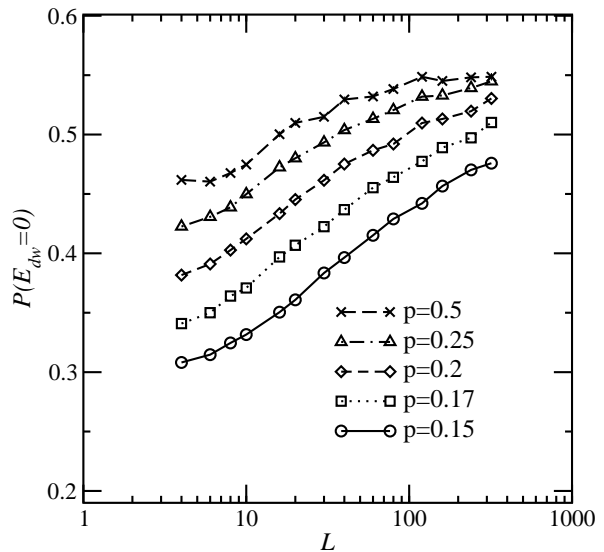}
\caption{\label{fig:Pstiff0} The probability $P(E_{dw}=0)$ for a
zero-energy domain wall as function of system size for selected
values of $p$. All error bars are at most of the symbol size. Lines
are guides to the eye.}
\end{figure}

Due to the quantization of the possible DW energies, a value
$\langle |E_{dw}| \rangle$ close to zero means that many system
exhibit actual zero DW energy $E_{dw} = 0$. We have also studied the
finite-size dependence of the probability $P(E_{dw} = 0)$ for a
zero-energy DW as a function of the system size, as shown in Fig.
\ref{fig:Pstiff0}.  One observes that zero-energy domain walls are
very common. Their probability of occurrence increases with the
concentration $p$ of the antiferromagnetic bonds (until $p=0.5$ due
to the symmetry of the model in $p,(1-p)$) and with the system size
leading to a convergence to a limiting value for $L \to \infty$.

The simplest assumption for the convergence of $\langle |E_{dw}|
\rangle$ and $P(E_{dw} = 0)$, at given values of $p$, to their
limiting values for $L\to\infty$ is a power law
\begin{equation}
F^{\star}(R,p,L) =
A^{\star}(p,R)+B^{\star}(p,R)L^{-\omega^\star(p,R)}\,.
\label{eq:simpleFit}
\end{equation}
The exponent $\omega^{\star}$ describes the rate of convergence.  It
depends on the fraction $p$ of the antiferromagnetic bonds and on
the aspect ratio $R$, as well as the constants $A^{\star}$ and
$B^{\star}$.  We have included the dependence on $R$ because below
we also study values of $R<1$, but for the moment we stick to $R=1$.

\begin{figure}
\includegraphics[width=0.9\columnwidth]{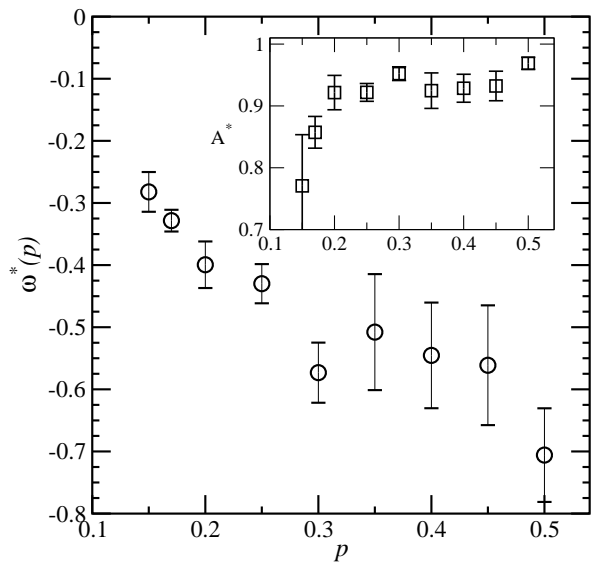}
\caption{\label{fig:EdwSimple} The results of fitting
Eqn.~(\ref{eq:simpleFit}) to the data ($R=1$): exponent
$\omega^{\star}$ (main plot) and limiting value $A^{\star}$
(inset).}
\end{figure}

When fitting Eqn.~(\ref{eq:simpleFit}) to the DW energy data for
$p\ge 0.15$, {\it i.e.} far enough away from $p_c$, and for sizes
$L\ge 10$, we obtain results as shown in Fig.\ \ref{fig:EdwSimple}.
For small concentrations $p$, the exponent $\omega^{\star}$ is less
negative than for $p$ close to $0.5$, corresponding to needing
larger sizes to reach the asymptotic value $A^{\star}$.  On the
other hand $A^{\star}$ itself seems to depend only weakly on $p$,
confirming the notion that everywhere inside the SG region the DW
energy reaches a non-zero plateau value. The dependence of
$\omega^{\star}$ appears like a strong violation of universality
inside the SG region. To investigate this assumption, whether it is
a true non-universality or just a finite-size scaling effect, we
have performed fits which also include corrections to scaling.  We
do this by fitting the data to a form
\begin{equation}
  F (R, p, L) = A (R, p) + \sum_{m = 1}^{m^*} B_m(R, p) L^{- m \omega}  \, .
\label{eq:fitFull}
\end{equation}
$F ( R , p, L )$ represents any function of $P (E_{dw})$.  Two
examples that we chose to study were $\langle |E_{dw}| \rangle$ and
the probability $P (E_{dw} = 0)$.

The number of correction-to-scaling terms included, $m^*$, was
chosen according to the amount of data to be fit.  Here we have
tried 2 and 3.  Of course, the values of the coefficients $A$ and
$B_m$ which are found by the fits depend somewhat on the choice of
$m^*$. The computed statistical errors do not include any allowance
for systematic errors due to the choice of the form of the scaling
function.  Fortunately, it turned out that the value of $A$ is
relatively insensitive to the choice of $m^*$.

In general, whether we chose $m^*$ to be 2 or 3, the computed $B_m$
coefficients did not all have the same sign.  This effect is caused
by the behavior at small $L$.  For this reason, there is no well
defined prescription for deciding what the best choice of the
exponent $\omega$ is. We have not included $\omega$ in the set of
free fitting parameters. Since it is the purpose of this section to
show that the behavior inside the full SG region is universal, we
have selected a value for $\omega$, such that the data for all
values of $p$ can be fit.  Thus the strong finite-size effects close
to $p_c$ can be explained also.  We have chosen the value $\omega =
2/3$, which is similar to the values used previously in other work
on this model,\cite{HY01,WHP03} empirically.  For a critical point
with $T_c > 0$, the general theory of finite-size
scaling\cite{Bar83} requires that $\omega = 1 / \nu$, where $\nu$ is
the correlation length exponent.  That result does not apply here,
however, since $T_c = 0$.

\begin{figure}[htb]
\includegraphics[width=3.4in]{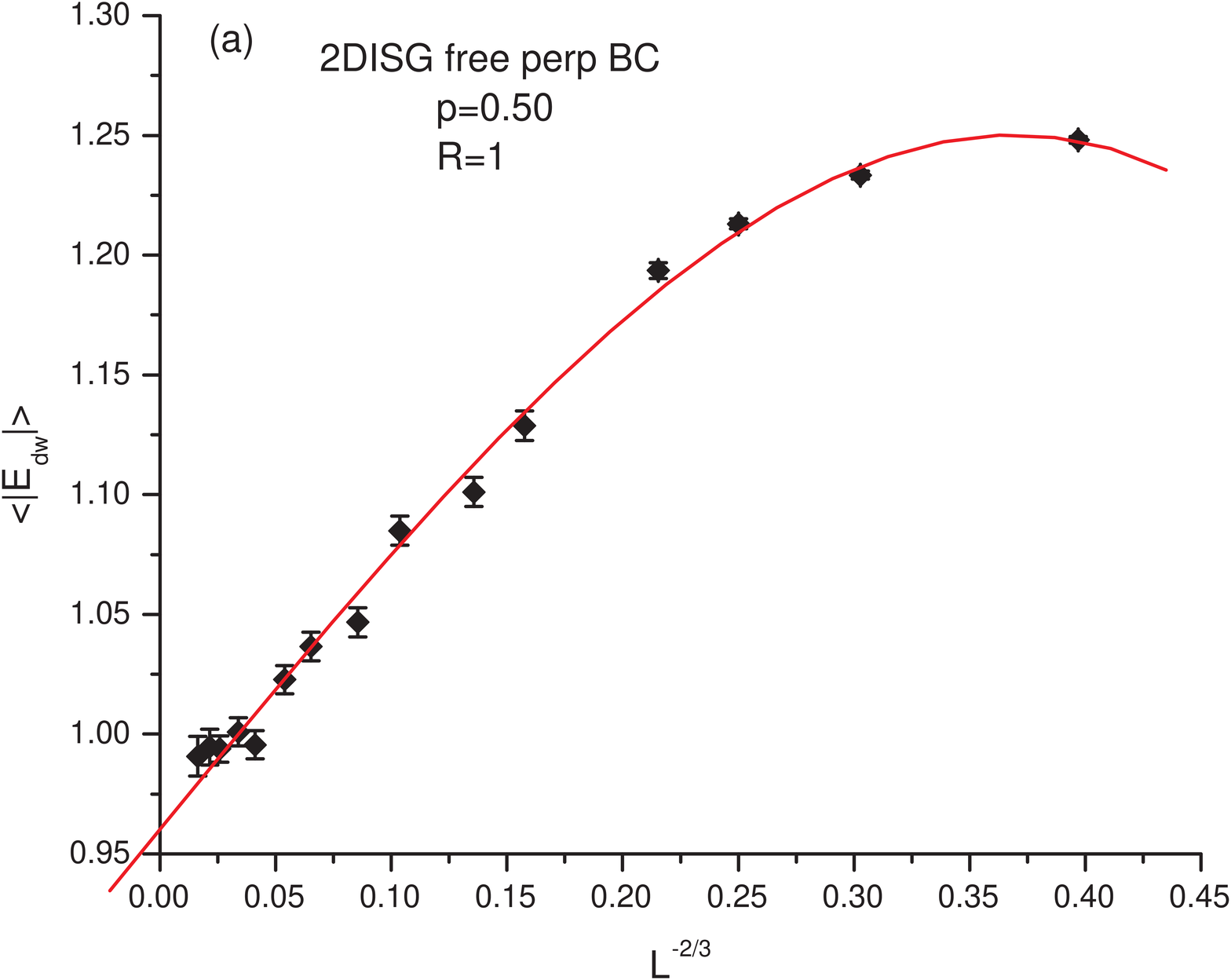}\quad
\includegraphics[width=3.4in]{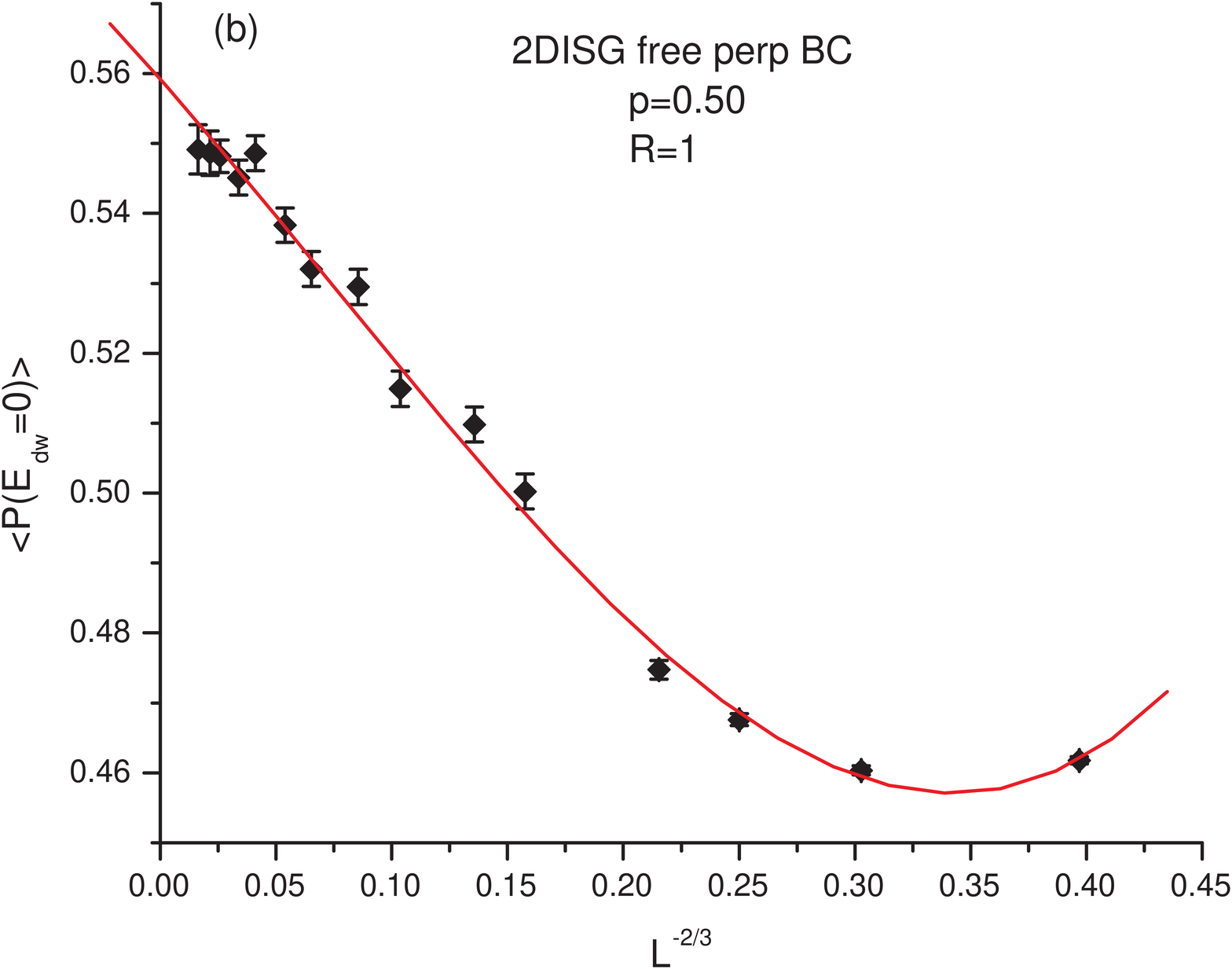}
\caption{\label{Fig.3}(color online) Finite-size scaling fits for
$p = 0.5$ and $R = 1$: (a) $\langle |E_{dw}| \rangle$ vs.
$L^{-2/3}$; (b) probability of $E_{dw} = 0$ vs. $L^{-2/3}$. The
error bars show one standard deviation.}
\end{figure}

As an example, our finite-size scaling fits (using $m^*=3$) to the
data using Eqn.~(\ref{eq:fitFull}) for the case $p = 0.5$ and $R =
1$ are shown in Fig.\ \ref{Fig.3}, the data for all system sizes was
included in the fits.  In the figures, we show $F(R,L)$ as a
function of $L^{-2/3}$. One observes a straight line for $L^{-2/3}
\to 0$, showing that the choice of the exponent is compatible with
the data. Note that the same $R = 1$ data was originally\cite{HY01}
fit (only for the case $p=0.5$) by making different assumptions
which are not consistent with Eqn.~(\ref{eq:fitFull}).

\begin{figure}[ht]
\includegraphics[width=3.4in]{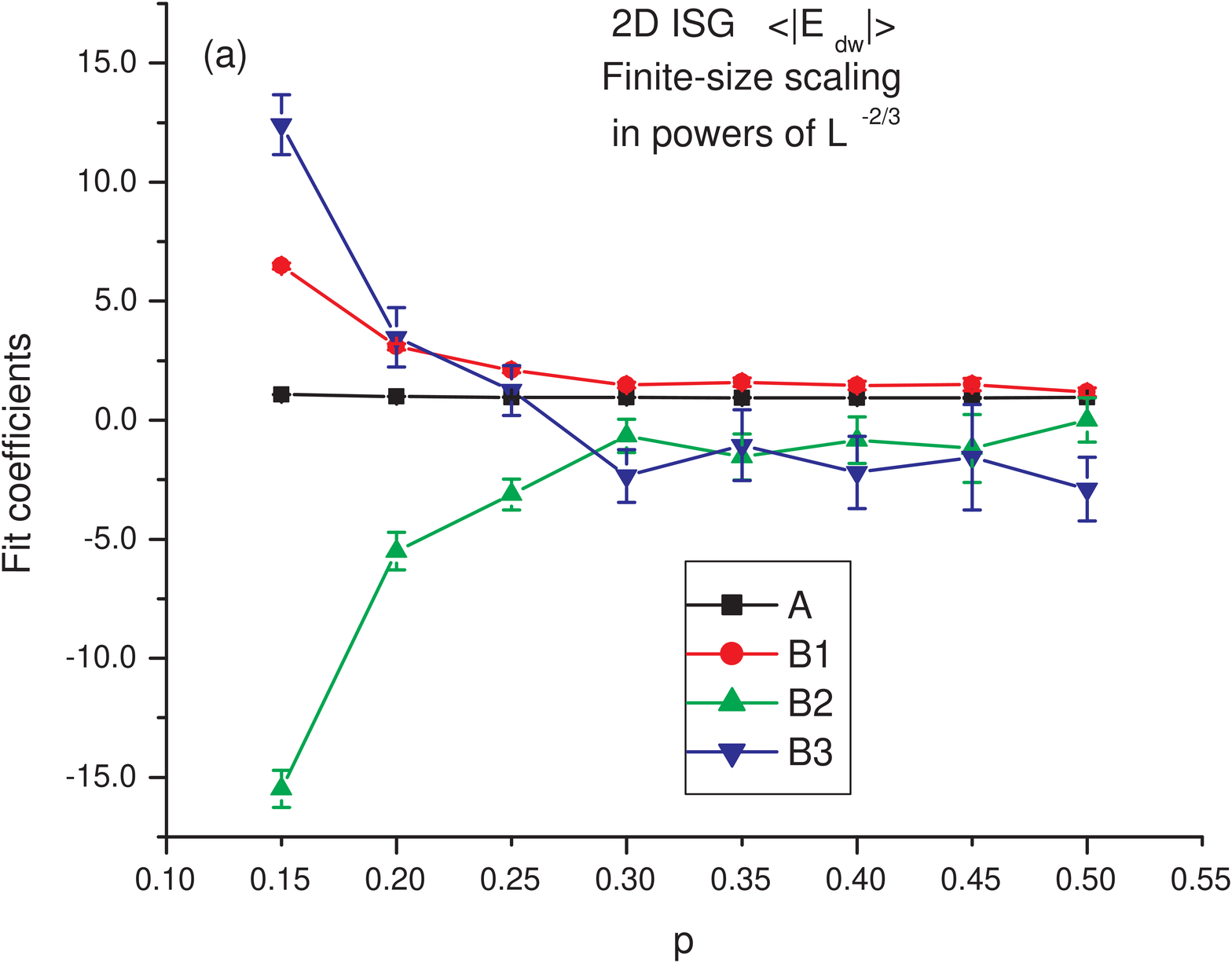}
\includegraphics[width=3.4in]{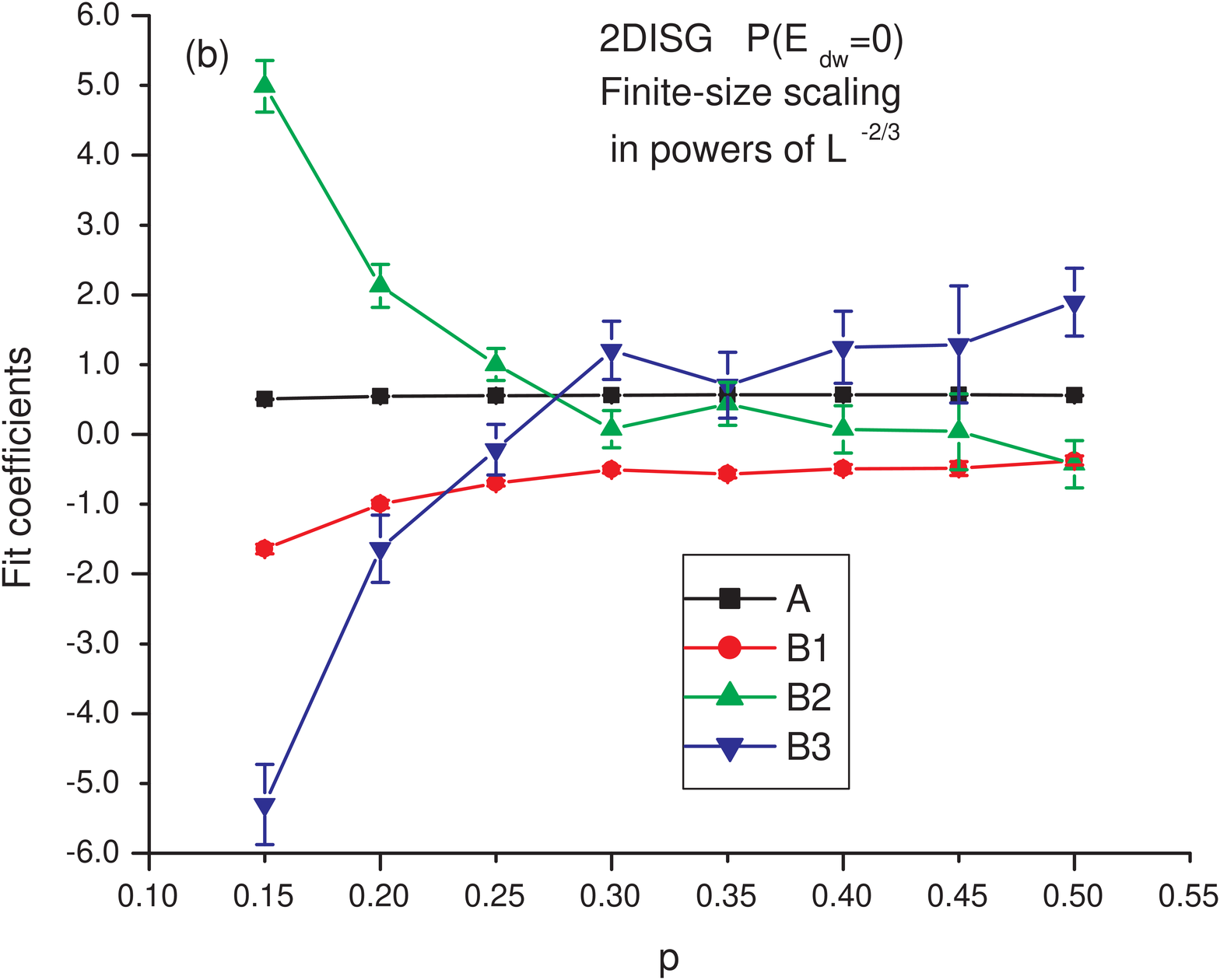}
\caption{\label{Fig.7}(color online) Fit coefficients (defined by
Eqn.\ \ref{eq:fitFull}) for $R = 1$: (a) $\langle | E_{dw} | \rangle$ vs. $p$;
(b) $P (E_{dw} = 0)$ vs. $p$.}
\end{figure}

Fig.\ \ref{Fig.7}(a) shows the fit coefficients for $\langle |
E_{dw} | \rangle$, as defined by Eqn.~(\ref{eq:fitFull}), as a
function of $p$ for $R = 1$.  The quality of fit was good
everywhere, proving that the apparent non-universality visible in
the dependence of the effective exponent $\omega^{\star}$ is due to
the presence of corrections to scaling. We again see that the value
of $A$, the estimate for $L \to \infty$, changes very little over
this range of $p$; the values are for $p\ge 0.3$ very similar to the
fit according to Eqn.~(\ref{eq:simpleFit}), but they are larger in
the present fit for small concentrations, $p< 0.3$. We also note
that there are substantial variations in the $B_m$ coefficients. In
particular, $B_3$ alternates in sign for the smaller values of $p$.
In Fig.\ \ref{Fig.7}(b) the results of fitting $P (E_{dw} = 0)$,
which shows the same type of behavior, are given. In this case the
signs of $B_2$ and $B_3$ alternate.

\subsection{Aspect-ratio scaling}

\begin{figure}[htb]
\includegraphics[width=0.8\columnwidth]{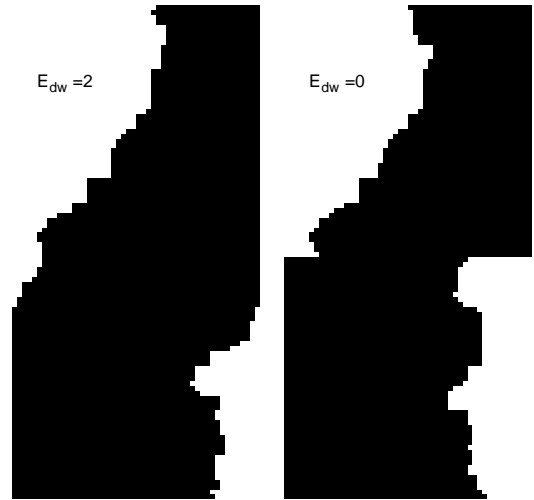}
\caption{\label{fig:normalDestroyed} Domain walls of subsystems are
not independent: Comparison of a minimum-energy DW for a system of
size $100\times 200$ (left) having $E_{dw}=2$ and of the same system
cut into two systems of size $100\times 100$ (right), having
$E_{dw}=0$. Note that in the left figure, the DW wraps around the
system in the $x$-direction. Note also that the DWs are highly
degenerate and the algorithm does not find DWs in a controlled way.
A ``jump'' in the DW at the right is visible. This figure
illustrates what a typical DW looks like.  The important point here
is the difference in the DW energies.}
\end{figure}

The main assumption when using AR scaling Eqn.~(\ref{eq:AR}) in the
limit $R\to 0$ is that different subsystems of size $M\times M$ are
independent of each other, which leads to a Gaussian distribution of
the DW energies $E_{dw}$ in the limit $R\to 0$. Nevertheless, one
can easily imagine that the DWs in different parts of the system are
not truly independent of each other. This is demonstrated for a
sample system in Fig.\ \ref{fig:normalDestroyed}, where the DW
energy of each subsystem (1)/(2) of size $100\times 100$ is
$E_{dw}^{(1)/(2)}=0$, but the DW energy of the full system is
$E_{dw}=2$. Also, in a previous study \cite{HBCMY02}, the validity
of AR scaling could not be established for the $\pm J$ model. Here,
we will show that AR scaling indeed works also for this case, {\it
i.e.} the distribution of the DW energies becomes indeed Gaussian,
despite the non-independencies of the domain walls inside the
different subsystems. Because the shape of the $P (E_{dw})$
distribution is changing with $R$, we expect that the AR scaling
will not be perfect in the range of our data.  As we shall see,
however, the deviations from the Gaussian fit become smaller than
our statistical errors for $R \le 1/4$.  Thus we are able to verify
that we are approaching the predicted scaling limit.

First, we concentrate on $p = 1/2$ where the distribution $P (
J_{ij} )$ of Eqn.\ (\ref{eq:PJ}) is symmetric about zero.  Thus in
this case the distribution $P (E_{dw})$ is (ignoring statistical
fluctuations) also symmetric around zero, for any values of $L$ and
$M$.  Therefore, we used $p = 1/2$ to collect data for sets of
random lattices with sequences of $L$ and $M$ having the aspect
ratios $R$ = 1, 1/2, 1/4, 1/8, 1/16 and 1/32.  For each value of
$R$, we then used finite-size scaling to extrapolate to the large
lattice limit.

\begin{figure}[htb]
\includegraphics[width=0.8\columnwidth]{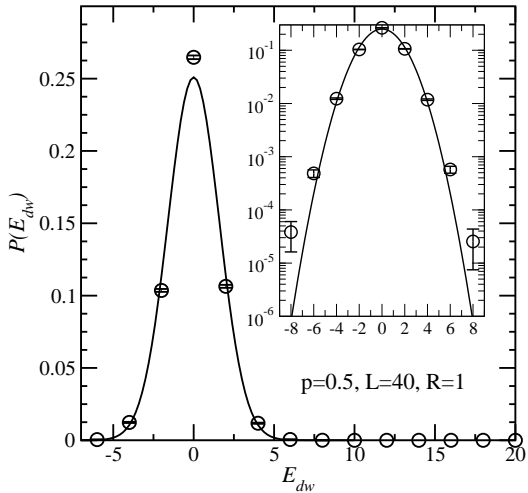}
\caption{\label{fig:gaussFitA} The symbols display the distribution
of DW energies for the case $L = M = 40$ ({\it i.e.} $R=1$) and
$p=0.5$, the lines shows the result of a fit to a Gaussian. Strong
deviations from a Gaussian distribution are visible at $P(E_{dw}=0)$
(main plot) and in the ``tail'' of the distribution (inset).}
\end{figure}

\begin{figure}[htb]
\includegraphics[width=0.8\columnwidth]{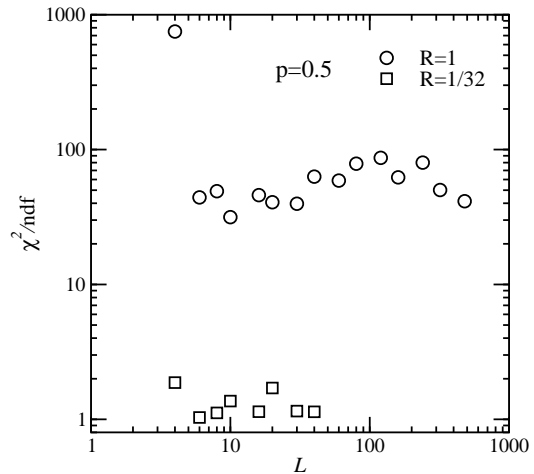}
\caption{\label{fig:ndf} Quality of the Gaussian fit for the DW
energy distribution measured by $\chi^2/$ndf as a function of the
block size $L$ for $p=0.5$ (circles: $R=1$, squares: $R=1/32$).}
\end{figure}

We start by studying $L \times L$ samples ($R=1$), to establish that
in this case the distribution of the DW energies is indeed not
Gaussian.  (Otherwise, it would not be surprising if it is also
Gaussian for $R \ll 1$.)  The distribution for $L=40$ is shown in
Fig.\ \ref{fig:gaussFitA}.  This distribution is obtained from an
average over 39000 realizations.  We have fit\cite{gnuplot} a
Gaussian to the data.  (Note that the fitting procedure ignores the
fact that the energies are quantized.)  Strong deviations from
Gaussian behavior are visible at $P(E_{dw}=0)$ (main plot) and in
the ``tail'' of the distribution (inset).  Correspondingly the
quality of the fit, {\it i.e.} $\chi^2$ per degree of freedom, as
given by our fitting program\cite{gnuplot} in this case, is very
high: $\chi^2/$ndf$=78$.  Even larger deviations are expected in the
tail.  To observe these, more sophisticated techniques would be
needed \cite{align,1dchain}, which is beyond the scope of this work.
The strong deviation from Gaussian behavior is not a finite-size
effect, as may be seen from Fig.\ \ref{fig:ndf}, where the circles
display $\chi^2/$ndf as a function of $L$ for the $R=1$ case.

The reader should also note that we find no evidence for any
"sawtooth" structure in the data. This is in marked contrast to the
case of periodic or antiperiodic boundary conditions in the {\bf y}
direction, which causes alternating values of $E_{dw}$ to have zero
probability. Therefore $P (E_{dw})$ does not become independent of
the boundary conditions even in the limit of large lattices.  This
is an aspect of the topological long-range order\cite{WHP03} which
exists in this model when the boundary conditions are periodic along
both {\bf x} and {\bf y}.

\begin{figure}[htb]
\includegraphics[width=0.8\columnwidth]{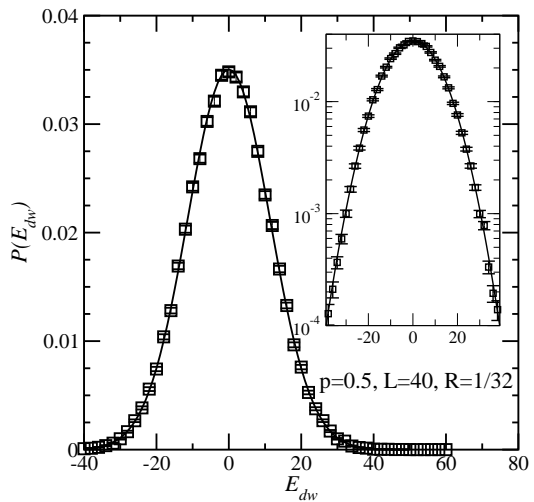}
\caption{\label{fig:gaussFitB} The symbols show the distribution of
DW energies for the case $L=40, M=1280$ ({\it i.e.} $R=1/32$) and
$p=0.5$, the lines shows the result of fitting to a Gaussian. Main
plot: linear ordinate, inset: logarithmic ordinate scale. The data
match the Gaussian very well.}
\end{figure}

In Fig.\ \ref{fig:gaussFitB} we show the probability distribution
$P(E_{dw})$ for a set of lattices with $p = 1/2$, $R = 1/32$ and $L
= 40$.  In this case, there are 90,000 random samples in the data
set.  Again, the distribution was fit to a Gaussian \cite{gnuplot}.
This fit is quite good, as demonstrated by the fact that the value
of $\chi^2/$ndf are is close to one.  Note that the precise value of
$\chi^2/$ndf depends on the number of points in the tail of the
distribution that are included in the fit, which is somewhat
arbitrary. The Gaussian behavior here is not a finite-size effect
either.  This can be seen from Fig.\ \ref{fig:ndf}, where the square
symbols display $\chi^2/$ndf as function of $L$ for the $R = 1/32$
case.

\begin{figure}
\includegraphics[width=3.4in]{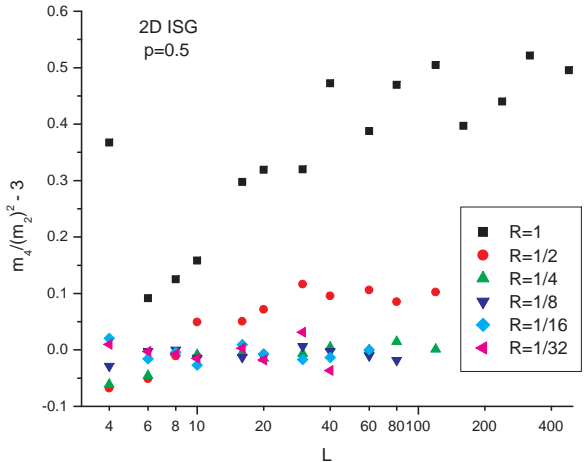}
\caption{\label{Fig.4}(color online) Kurtosis vs. $L$  ($p =
0.5$.) The {\bf x} axis is scaled logarithmically.}
\end{figure}

When $p = 0.5$ the configuration averages of all the odd moments of
$P (E_{dw})$ must vanish.  Then the dominant contribution to
non-Gaussian behavior will be the kurtosis, which simplifies to $m_4
/ (m_2)^2 - 3$, where $m_i$ is the $i^{th}$ moment of $P (E_{dw})$,
under these conditions.  Fig.\ \ref{Fig.4} shows the behavior of the
kurtosis of the $P (E_{dw})$ distributions at $p = 0.5$ as a
function of $L$ and $R$. It shows that for $R \le 1/4$ the kurtosis
is negligible, except when $L < 8$, quantifying the convergence
towards a Gaussian distribution for small $R$.

This convergence can be also seen when fitting $\langle |E_{dw}|
\rangle$ to Eqn.~(\ref{eq:fitFull}), now for aspect ratios $R < 1$.
In Table I we show the results of the finite-size scaling fits at $p
= 0.5$.  According to Eqn.~(\ref{eq:AR}), the prediction of AR
scaling theory is that the ratio $A ( R ) / A ( 2 R )$ should
approach $\sqrt 2 \approx 1.414$ for small $R$. Considering the
statistical errors and the corrections to scaling, the agreement is
quite satisfactory. Here we observe that for $R < 1/4$ the scaling
assumption is reasonably accurate.

\begin{table}
\caption{\label{tab:table1}$\langle |E_{dw}| \rangle$ for $p = 0.5$,
extrapolated to $L = \infty$, as a function of the aspect ratio,
$R$.  The $m^*$ and $A$ are defined by Eqn.~(\ref{eq:fitFull}), and
the $eA$ are statistical error estimates.}
\begin{ruledtabular}
\begin{tabular}{lcccc}
 $R$ & $m^*$ & $A$ & $eA$ & $A ( R ) / A ( 2 R )$ \\
\hline
 1.0    & 3 & 0.96025 & 0.00792 & \\
 0.5    & 2 & 1.81967 & 0.00471 & 1.895 \\
 0.25   & 2 & 2.89085 & 0.00671 & 1.589 \\
 0.125  & 2 & 4.28859 & 0.01013 & 1.484 \\
 0.0625 & 2 & 6.21247 & 0.02387 & 1.449 \\
 0.03125 & 1 & 8.98954 & 0.03402 & 1.447 \\
\end{tabular}
\end{ruledtabular}
\end{table}

Table II shows the results of the finite-size scaling fits for the
probability that $E_{dw} = 0$ at $p = 0.5$.  Since, from the central
limit theorem, for the sum $\Sigma$ of $K$ suitably iid integers,
$P(\Sigma=0) \sim K^{-0.5}$, via $K \sim R^{-1}$ the prediction of
AR scaling theory for this quantity is that the ratio $A ( R ) / A (
2 R )$ should approach $1 / \sqrt 2 \approx 0.707$ for small $R$.
Again, for $R < 1/4$ the agreement is about as good as could be
expected.

\begin{table}
\caption{\label{tab:table2}Probability of $E_{dw} = 0$ for $p =
0.5$, extrapolated to $L = \infty$, as a function of the aspect
ratio, $R$. The $m^*$ and $A$ are defined by
Eqn.~(\ref{eq:fitFull}), and the $eA$ are statistical error
estimates.}
\begin{ruledtabular}
\begin{tabular}{lcccc}
 $R$ & $m^*$ & $A$ & $eA$ & $A ( R ) / A ( 2 R )$ \\
\hline
 1.0    & 3 & 0.55921 & 0.00294 & \\
 0.5    & 3 & 0.33215 & 0.00261 & 0.594 \\
 0.25   & 2 & 0.21619 & 0.00135 & 0.651 \\
 0.125  & 2 & 0.14736 & 0.00139 & 0.682 \\
 0.0625 & 2 & 0.10264 & 0.00144 & 0.697 \\
 0.03125 & 1 & 0.07053 & 0.00069 & 0.687 \\
\end{tabular}
\end{ruledtabular}
\end{table}

We have also studied the AR approach for $p=0.15$ and found similar
results (not shown). For aspect ratios $R < 1/4$ the assumptions of
a Gaussian distribution of the DW energies is again valid. Hence, we
believe that inside the entire SG region the behavior is consistent
with the assumptions of the AR approach.

\subsection{Discussion of AR scaling}

At the end of Section II we argued that for small $R$ the system may
be thought of as a set of $L \times L$ subsystems stacked in the
{\bf y} direction.  This implies that the behavior of $E_{dw}$ in
this limit should obey Eqn.~(\ref{eq:AR}) with $\theta = 0$.  As we
have seen, our numerical results are indeed consistent with this
argument.  However, the hand-waving argument does not specify how
the subsystems are to be connected to each other.  Before actually
seeing the numerical results, we were far from confident that this
argument would correctly predict what was found.

There is another apparently reasonable argument, which leads one to
expect that the results for small $R$ should not give this answer.
As $R \to 0$, we know that the width of the $E_{dw}$ distribution
will diverge.  This means that for most of our sample lattices $|
E_{dw} | \gg 1$ in this limit.  One might have thought, therefore,
that the fact that $E_{dw}$ is quantized would no longer matter.
Thus one might have guessed that the system would behave in this
limit as if $\theta \approx -0.28$, the value for an unquantized
distribution.

Therefore, our results indicate that a Bohr correspondence
principle, that in the limit of large quantum numbers the results
should be indistinguishable from those of an unquantized system,
does not apply.  How can we understand this?  In our opinion, what
we learn from this is that $\theta = 0$ is directly connected to
the existence of topological long-range order\cite{WHP03} at $T =
0$.

One might try to argue that $R = 1 / 32$ is not small enough, and
that for even smaller $R$ one would indeed find a crossover to
$\theta \approx -0.28$.  Such a viewpoint is suggested by the recent
work of J\"{o}rg {\it et al.}\cite{JLMM06}  On the other hand a more
detailed analysis\cite{Fis06} of the low temperature behavior of the
specific heat of the $\pm J$ model does not support this
interpretation.

\section{Summary}

We have studied the statistics and finite-size scaling behavior of
GS and DW energies for the 2D Ising random-bond spin glass with an
mixture of $+1$ and $-1$ bonds, where $p$ is the concentration of
antiferromagnetic bonds. By using sophisticated matching algorithms,
we can calculate the GS energy and DW energies for quite large
systems exactly, which allows us to study the scaling behavior very
precisely. We find that the scaling behavior of the GS energy can be
described everywhere inside the SG region by the same simple scaling
function Eqn.~(\ref{eq:e0L}), as predicted by Ref.\
\onlinecite{CHK04}. Furthermore, when looking at the fluctuations of
the GS energy, we find, after carefully taking into account
finite-size corrections, that they follow the predicted simple
$L^{-1}$-scaling with an amplitude which has a cusp singularity at
the ferromagnetic-SG transition $p_c$. The singularity is described
by an exponent which is close to $2/3$.

The behavior of the DW energy is characterized close to $p_c$, by
huge finite-size effects. Nevertheless, the results are compatible
with a convergence of $\langle |E_{dw}| \rangle $ to a finite
plateau value everywhere inside the SG phase. This convergence can
be described, again universally inside the SG phase, by a single
exponent $\omega \approx 2/3$, just by taking into account
higher-order corrections to scaling.  Finally, we have studied $L
\times M$ rectangular lattices with aspect ratio $R$ between 1/32
and 1. We have demonstrated by an example that one assumption
underlying AR scaling, i.e.\ the assumed independence of the DW
energies of different blocks, is not strictly valid. Nevertheless,
we find that for large lattices the probability distribution of
$E_{dw}$ in the SG region of the phase diagram approaches for $R \le
1 / 4$ a Gaussian centered at $E_{dw} = 0$. Hence, in the small $R$
limit the behavior obeys the AR scaling predictions of Carter, Bray
and Moore.\cite{CBM02}

\begin{acknowledgments}
RF is grateful to S. L. Sondhi, F. D. M. Haldane and D. A. Huse
for helpful discussions, and to Princeton University for providing
use of facilities.  AKH acknowledges financial support from the
{\it VolkswagenStiftung} (Germany) within the program
``Nachwuchsgruppen an Universit\"{a}ten'' and from the DYGLAGEMEN
program funded by the EU.

\end{acknowledgments}



\end{document}